\documentclass[a4paper]{jpconf}
\usepackage{graphicx,hyperref}

\begin{document}
\title{Landau quantization and neutron emissions by nuclei in the crust of a magnetar}

\author{N Chamel$^1$, Y D  Mutafchieva$^2$, Zh K Stoyanov$^2$, L M Mihailov$^3$ and R L Pavlov$^2$}

\address{$^1$ Institute of Astronomy and Astrophysics, Universit\'e Libre de Bruxelles, CP 226, Boulevard du Triomphe, B-1050 Brussels, Belgium}
\address{$^2$ Institute for Nuclear Research and Nuclear Energy, Bulgarian Academy of Sciences, 72 Tsarigradsko Chaussee, 1784 Sofia, Bulgaria}
\address{$^3$ Institute of Solid State Physics, Bulgarian Academy of Sciences, 72 Tsarigradsko Chaussee, 1784 Sofia, Bulgaria}

\ead{nchamel@ulb.ac.be}

\begin{abstract}
Magnetars are neutron stars endowed with surface magnetic fields of the order of $10^{14}-10^{15}$~G, and with presumably much 
stronger fields in their interior. 
As a result of Landau quantization of electron motion, the neutron-drip transition in the crust of a magnetar is shifted to 
either higher or lower densities depending on the magnetic field strength. The impact of nuclear uncertainties is explored 
considering the recent series of Brussels-Montreal microscopic nuclear mass models. All these models are based on the Hartree-Fock-Bogoliubov 
method with generalized Skyrme functionals. They differ 
in their predictions for the symmetry energy coefficient at saturation, and for the stiffness of the neutron-matter equation 
of state. For comparison, we have also considered the very accurate but more phenomenological model of Duflo and Zuker. Although the equilibrium 
composition of the crust of a magnetar and the onset of neutron emission are found to be model dependent, the quantum oscillations of the threshold 
density are essentially universal. 
\end{abstract}

\section{Introduction}

At the end point of stellar evolution, neutron stars are not only the most compact stars in the universe, but also the strongest magnets~\cite{hae07}. 
In particular, magnetic fields of the order of $10^{14}-10^{15}$~G have been measured at the surface of soft gamma-ray repeaters and anomalous x-ray 
pulsars~\cite{mcgill14}, thus dubbed \emph{magnetars}. On the other hand, numerical simulations have shown that the internal magnetic field could be 
even stronger, up to about $10^{18}$~G~\cite{deba15}. The outermost layer of a neutron star is thought to consist of a solid crust, whose atoms are 
fully ionized by the gravitational pressure (for a review, see Ref.~\cite{lrr}). With increasing depth, nuclei become progressively more neutron rich 
by capturing electrons until at some point, neutrons start to drip out of nuclei. The presence of a neutron liquid in the crust of a magnetar is expected 
to leave its imprint on various observed astrophysical phenomena like sudden spin-ups~\cite{dib08,alpar14} and spin-downs~\cite{archi13,mus14,duncan13,kantor14} 
(generally referred to as ``glitches'' and ``anti-glitches'' respectively), quasiperiodic oscillations detected in the giant flares from soft gamma-ray 
repeaters~\cite{and09,chamel2013,pass14} and cooling~\cite{agui09}. 

We have recently studied the effects of a strong magnetic field on the neutron-drip transition in the crust of a magnetar~\cite{chamel2012,chamel2015b}. We have shown 
that the neutron-drip density and pressure increase almost linearly with the magnetic field strength in the strongly quantizing regime. In the weakly quantizing 
regime, the variations of the neutron-drip density with magnetic field strength exhibit typical quantum oscillations. The neutron-drip transition in a magnetar, 
as compared to unmagnetized neutron stars, can thus be shifted to either higher or lower densities depending on the magnetic field strength. 
In this paper, we explore the role of nuclear uncertainties on the neutron-drip transition considering different nuclear mass models.

\section{Neutron emission in dense magnetized matter} 
\label{model}

During the gravitational collapse of the core of massive stars (whose mass lies in the range $\sim 8-10~M_\odot$, where 
$M_\odot$ is the mass of the Sun), it is generally assumed that all kinds of reactions can occur so that the hot dense matter 
constituting the newly born neutron star remains in thermodynamic equilibrium. Eventually the neutron star interior becomes 
cold and fully ``catalyzed''~\cite{hae07}. Although the temperature inside mature neutron stars with an age $\sim 10^4-10^5$ years 
is typically of the order of $10^7-10^8$~K, matter is so highly degenerate that for most practical purposes the temperature 
can be set to zero. Since the pressure inside the star must vary continuously, and since the temperature is fixed, the suitable 
thermodynamic potential for determining the internal composition of a neutron star is the Gibbs free energy per nucleon $g$ (this 
still remains the case even in the presence of a strong magnetic field, as shown in Ref.~\cite{chamel2015a}). In the deep region 
of the outer crust on which we focus, atoms are fully ionized by the pressure, and electrons can be treated as an almost ideal 
Fermi gas. The main correction arises from the electrostatic lattice energy (for a discussion of various other corrections, see e.g. 
Ref.~\cite{chamel2015a}). In each layer at pressure $P$, the crust is assumed to be made of only one type of ions with proton number 
$Z$ and mass number $A$, arranged on a body-centred cubic lattice. Due to the presence of a strong magnetic field $B$, the electron 
motion perpendicular to the field is quantized into Landau levels. The magnetic field $B_\star=B/B_{\rm crit}$ can be conveniently 
measured in units of the critical magnetic field defined by 
\begin{equation}
\label{eq:Bcrit}
B_{\rm crit}=\left(\frac{m_e c^2}{\alpha \lambda_e^3}\right)^{1/2}\simeq 4.4\times 10^{13}\, \rm G\, ,
\end{equation}
where $m_e$ is the electron mass, $c$ the speed of light, $\alpha$ is the fine structure constant, and $\lambda_e$ is the electron 
Compton wavelength. Typically, Landau quantization effects on the composition of the crust are negligible unless $B_\star \gg 1$. 
Further details on our crust model can be found in Refs.~\cite{chamel2012,chamel2015b}. 

In any region of the outer crust, the pressure is provided by the degenerate electron gas so that the electron chemical potential 
$\mu_e$ increases with depth until it reaches the threshold for neutron drip~\cite{chamel2015b} 
\begin{equation}\label{eq:n-drip-mue}
\mu_e(n_e,B_\star) + \frac{4}{3}C e^2 n_e^{1/3} Z^{2/3} =  \mu_e^{\rm drip}(A,Z)\, ,
\end{equation}
where $n_e$ is the electron number density, $e$ is the proton electric charge, $C\approx -1.444$ is the body-centred cubic lattice structure 
constant, and 
\begin{equation}\label{eq:muedrip}
\mu_e^{\rm drip}(A,Z)\equiv \frac{-M(A,Z)c^2+A m_n c^2}{Z} +m_e c^2 \, ,
\end{equation}
with $M(A,Z)$ the nuclear mass (including the rest mass of $Z$ protons, $A-Z$ neutrons and $Z$ electrons \footnote{The reason for 
including the electron rest mass in $M(A,Z)$ is that experimental \emph{atomic} masses are generally tabulated rather than \emph{nuclear} masses.}), 
and $m_n$ the neutron mass. The equilibrium nucleus is found by minimizing the Gibbs free energy per nucleon. As discussed in Ref.~\cite{chamel2015c}, 
this nucleus is actually stable against neutron emission, but unstable against electron captures accompanied by neutron emission. 
Equation~(\ref{eq:n-drip-mue}) must be generally solved numerically. Nevertheless, an analytical solution can be obtained in the strongly 
quantizing regime for which electrons lie in the lowest Landau level and are ultrarelativistic ($\mu_e\gg m_e c^2$). In such case, 
the neutron-drip density and pressure are approximately given by~\cite{chamel2015b}
\begin{equation}\label{eq:rhodripB}
n_{\rm drip}\approx \frac{A}{Z} \frac{B_\star \mu_e^{\rm drip}(A,Z)}{ 2\pi^2 \lambda_e^3 m_e c^2}
 \biggl[1-\frac{4}{3} C \alpha Z^{2/3}\left(\frac{B_\star}{2\pi^2}\right)^{1/3}\left(\frac{m_e c^2}{\mu_e^{\rm drip}(A,Z)}\right)^{2/3}\biggr]\, ,
\end{equation}
\begin{equation}\label{eq:PdripB}
P_{\rm drip}\approx \frac{B_\star \mu_e^{\rm drip}(A,Z)^2}{4 \pi^2 \lambda_e^3 m_e c^2}\biggl[1-\frac{1}{3} C \alpha Z^{2/3}
\left(\frac{4 B_\star }{\pi^2}\right)^{1/3} \left(\frac{m_e c^2}{\mu_e^{\rm drip}(A,Z)}\right)^{2/3}   \biggr]\, ,
\end{equation}

For any given magnetic field, the composition of the outer crust and the onset of neutron drip are completely determined by nuclear masses. 
The nuclei expected to be found in the bottom layers of the outer crust are so neutron rich that their masses have not yet been measured. 
For this reason, nuclear models must be employed. We have made use of the recent series of Brussels-Montreal microscopic nuclear 
mass models~\cite{gcp13a,gcp13b}. These models are based on the self-consistent Hartree-Fock-Bogoliubov (HFB) method using generalized Skyrme 
functionals (see e.g. Ref.~\cite{chamel2015d} for a review of the latest Brussels-Montreal models). These models were primarily fitted 
to the 2353 measured masses of nuclei with $N$ and $Z \geq 8$ from the 2012 Atomic Mass Evaluation~\cite{audi2012}, with a root-mean square 
deviation $\sim 0.5-0.6$~MeV. At the same time, the underlying BSk functionals were constrained to reproduce various properties of 
infinite homogeneous nuclear matter, most of which are summarized in Table~\ref{tab1}. These parameters are defined by first writing the energy per 
nucleon of nuclear matter of density $n$ and charge asymmetry $\eta = (n_n - n_p)/n$ ($n_n$ and $n_p$ are the neutron and proton densities respectively) 
in the form
\begin{equation}
e(n,\eta)\approx e(n,\eta=0)+ e_{\rm sym}(n)\eta^2\ ,
\end{equation}
in which the first term on the right-hand side is just the energy per nucleon of charge-symmetric nuclear matter; we have neglected charge-symmetry 
breaking terms, such as those arising from the neutron-proton mass difference. We then expand $e(n,\eta=0)$ and $e_{\rm sym}(n)$ about the equilibrium 
density $n_0$ in powers of $\epsilon = (n - n_0)/n_0$, thus
\begin{equation}
e(n,\eta=0) \approx a_v + \frac{1}{18}K_v\epsilon^2 \, ,
\end{equation}
and
\begin{equation}
e_{\rm sym}(n) \approx J + \frac{1}{3}L\epsilon + \frac{1}{18}K_{\rm sym}\epsilon^2\, .
\end{equation}
For pure neutron matter ($\eta=1$) at density $n_0$, the pressure is approximately given by $P_n\approx L n_0/3$. Therefore, $L$ is a measure of the 
stiffness of the neutron-matter equation of state. The functionals BSk22, BSk23, BSk24 and BSk25 were all fitted to the same realistic neutron-matter 
equation of state (as obtained from many-body calculations using realistic two- and three-nucleon forces), but were constrained to different symmetry energy 
coefficients, $J = 32$, $31$, $30$ and $29$~MeV, respectively. BSk26 was constrained to the same symmetry-energy coefficient $J = 30$~MeV as BSk24, 
but was fitted to a softer neutron-matter equation of state. BSk27$^*$ yields the softest possible neutron-matter equation of state consistent with 
recent quantum Monte Carlo calculations and with the same symmetry-energy coefficient. The model HFB-27$^*$ is also the most accurate of this series. 
The range of values of $J$ and $L$ spanned by these functionals are consistent with constraints coming from the combined analysis of various 
experiments~\cite{fantina2015} (see also the discussion in Sec.~IIIC in Ref.~\cite{gcp13a}). Moreover, the incompressibility coefficient $K_v$ falls in 
the empirical range $240\pm10$~MeV \cite{colo2004}. Finally, the high-density equation of state of symmetric nuclear matter predicted by these functionals 
are compatible with the constraints inferred from the analysis of heavy-ion collision experiments~\cite{dan02,lynch09}. 

\begin{table}[h]
\caption{\label{tab1}Nuclear-matter properties predicted by recent Brussels-Montreal nuclear energy density functionals~\cite{gcp13a,gcp13b}.}
\begin{center}
\begin{tabular}{lllllll}
\br
 & BSk22 & BSk23 & BSk24 & BSk25 & BSk26 & BSk27$^*$\\
\mr
$n_0$ [fm$^{-3}$] & 0.1578 & 0.1578 & 0.1578 & 0.1587 & 0.1589 & 0.1586\\
$a_v$ [MeV] & -16.088 & -16.068 & -16.048 & -16.032 & -16.064 & -16.051\\
$K_v$ & 245.9 & 245.7 & 245.5 & 236.0 & 240.8 & 241.6\\
$J$ [MeV] & 32.0 & 31.0 & 30.0 & 29.0 & 30.0 & 30.0 \\
$L$ [MeV] & 68.5 & 57.8 & 46.4 & 36.9 & 37.5 & 28.5 \\
$K_{\rm sym}$ [MeV] & 13.0 & -11.3 & -37.6 & -28.5 & -135.6 & -221.4 \\
\br
\end{tabular}
\end{center}
\end{table}

\begin{figure}
\begin{center}
\includegraphics[scale=0.5]{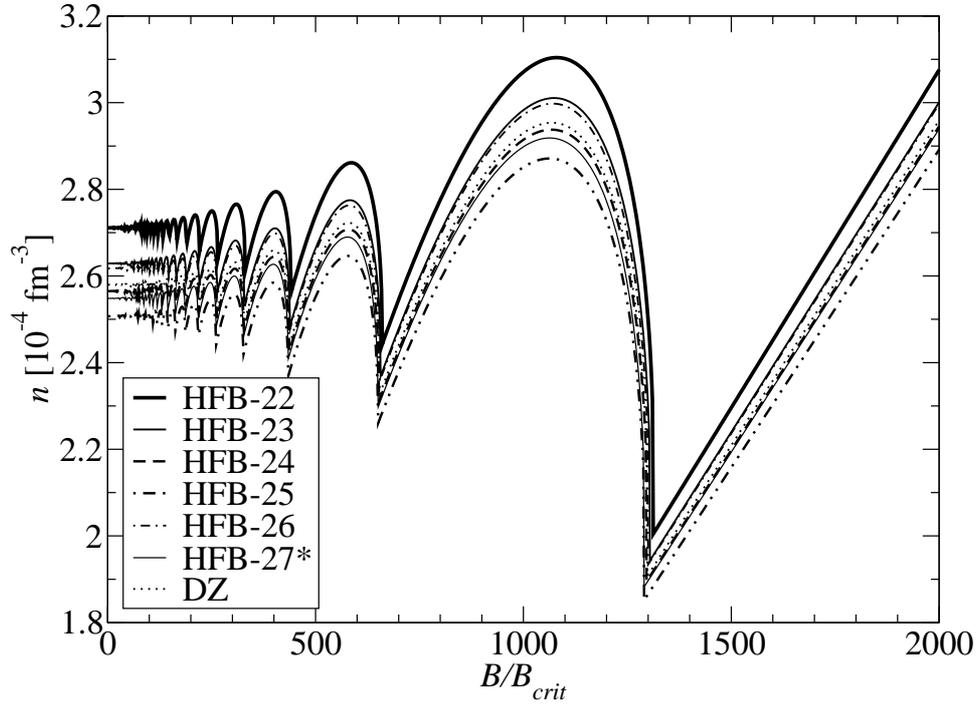}
\end{center}
\caption{Threshold baryon number density for the onset of neutron drip in the crust of a neutron star as a function of magnetic field strength, as predicted by 
different nuclear mass models. 
}
\label{fig1}
\end{figure}

\begin{figure}
\begin{center}
\includegraphics[scale=0.5]{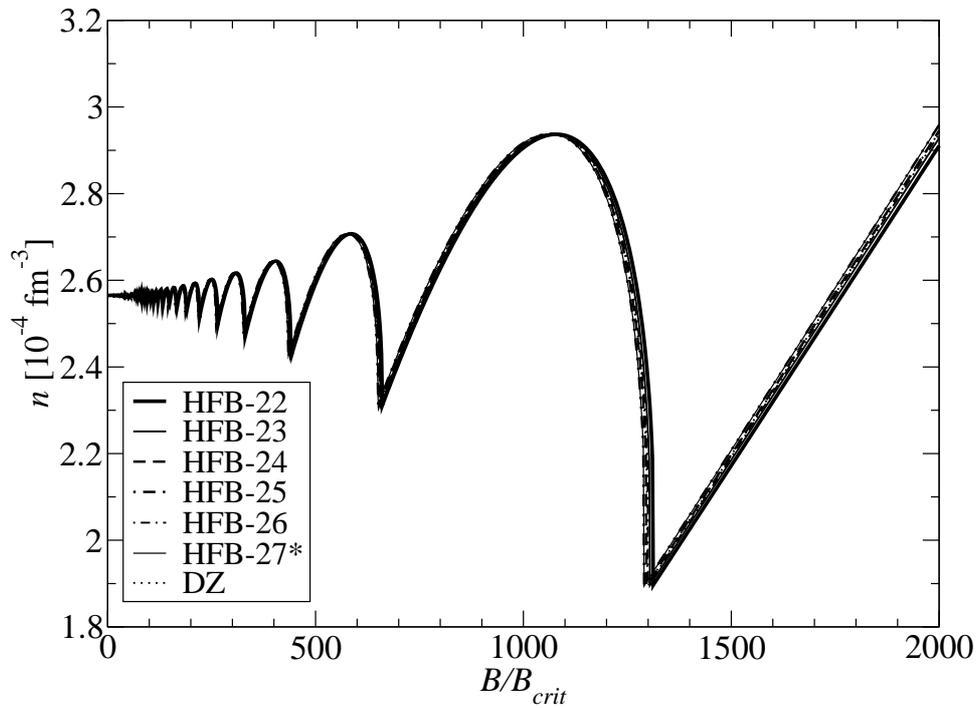}
\end{center}
\caption{Same as Figure~\ref{fig1} after a suitable rescaling of the threshold density to that of HFB-24. 
}
\label{fig2}
\end{figure}

Taking the values of the nuclear masses from BRUSLIB~\cite{bruslib}, we have determined numerically the equilibrium composition of the outer crust of a magnetar 
for different magnetic field strengths. For comparison, we have also considered the more phenomenological model of Duflo and Zuker~\cite{dz95}. In the 
search of the equilibrium nucleus, we have made use of mass models only when experimental measurements from the 2012 Atomic Mass Evaluation~\cite{audi2012}
were not available. The composition 
at the neutron-drip transition is indicated in Table~\ref{tab2}. For all models but HFB-22, the equilibrium nucleus is the same with or without the presence of 
a strong magnetic field. For model HFB-22, the equilibrium nucleus is found to be either $^{122}$Kr or $^{128}$Sr, depending on the magnetic field strength. The 
reason lies in the fact that the threshold electron chemical potentials $\mu_e^{\rm drip}$ for these two nuclides differ by less than $0.2\%$. As shown in 
Figs.~\ref{fig1} and \ref{fig2}, the oscillations of the neutron-drip density as a function of $B_\star$ are almost universal. In particular, in the presence of 
a nonquantizing magnetic field (i.e. more than one Landau level is populated), the shifts in the neutron-drip density are bounded, both from above and from below. 
The limits are approximately given by~\cite{chamel2015b}:  
\begin{equation}
\frac{n_{\rm drip}^{\rm min}}{n_{\rm drip}(B_\star=0)}\approx \frac{3}{4} \, ,
\end{equation} 
\begin{equation}
\frac{n_{\rm drip}^{\rm max}}{n_{\rm drip}(B_\star=0)}\approx \frac{1}{72}\left(35+13\sqrt{13}\right)\, .
\end{equation} 

\begin{table}[h]
\caption{\label{tab2}Equilibrium nucleus and electron chemical potential at the neutron-drip transition in the crust of a magnetar, 
as predicted by different nuclear mass models.}
\begin{center}
\begin{tabular}{lll}
\br
       & $^{A}$X & $\mu_e^{\rm drip}$ (MeV) \\
\mr
HFB-22 & $^{122}$Kr/$^{128}$Sr  & 24.97/25.01\\
HFB-23 & $^{126}$Sr & 24.88 \\
HFB-24 & $^{124}$Sr & 24.81 \\
HFB-25 & $^{122}$Sr & 24.76 \\
HFB-26 & $^{126}$Sr & 24.85 \\
HFB-27$^*$ & $^{124}$Sr & 24.76 \\
DZ         & $^{118}$Kr & 24.87 \\
\br
\end{tabular}
\end{center}
\end{table}

\section{Conclusions}

We have pursued our study of the effects of Landau quantization on the onset of neutron emission by nuclei in the crust of a magnetar~\cite{chamel2015b}. In 
particular, we have explored the impact of nuclear uncertainties considering the recent series of Brussels-Montreal Hartree-Fock-Bogoliubov 
nuclear mass models, from HFB-22 to HFB-27$^*$~\cite{gcp13a,gcp13b}. These models yield equally good fits to essentially all experimental masses, with a root-mean-square 
deviation of about $0.5-0.6$ MeV. However, they lead to different predictions of nuclear-matter properties thus reflecting the current lack of 
knowledge of the symmetry energy, and the stiffness of the neutron-matter equation of state. For comparison, we also considered the more accurate 
but also more phenomenological model of Duflo and Zuker~\cite{dz95}. Although the equilibrium nucleus at the bottom the outer crust, and the onset of neutron 
emission are found to be model dependent, the oscillations of the threshold density as a function of the magnetic field strength are almost universal. 
The role of the magnetic field on the neutron-drip density in magnetar crusts could thus be a priori inferred from the value of the neutron-drip density 
in unmagnetized neutron star crusts. On the other hand, the change of nuclear masses due to the strong magnetic field~\cite{bas2015}, which we 
have not taken into account, may undermine this universal behaviour. 

\ack 
This work was financially supported by Fonds de la Recherche Scientifique - FNRS (Belgium),  Wallonie-Bruxelles-International 
(Belgium), the Bulgarian Academy of Sciences, the Bulgarian National Science Fund under contract No. DFNI-T02/19, and the 
European Cooperation in Science and Technology (COST)  Action MP1304 ``NewCompStar''.

\end{document}